\documentclass[reprint,amsmath, amssymb, superscriptaddress]{revtex4-1}

\usepackage{graphicx}

\renewcommand{\thefigure}{\textbf{\arabic{figure}}}

\begin{document}

\title{Electric field spectroscopy of material defects in transmon qubits}

\author{J\"urgen Lisenfeld*}
\affiliation{Physikalisches Institut, Karlsruhe Institute of Technology, 76131 Karlsruhe, Germany}	
\email[]{e-Mail: juergen.lisenfeld@kit.edu}
\author{Alexander Bilmes}
\affiliation{Physikalisches Institut, Karlsruhe Institute of Technology, 76131 Karlsruhe, Germany}
\author{Anthony Megrant}
\affiliation{Google, Santa Barbara, California 93117, USA}
\author{Rami Barends}
\affiliation{Google, Santa Barbara, California 93117, USA}
\author{Julian Kelly}
\affiliation{Google, Santa Barbara, California 93117, USA}
\author{Paul Klimov}
\affiliation{Google, Santa Barbara, California 93117, USA}
\author{Georg Weiss}
\affiliation{Physikalisches Institut, Karlsruhe Institute of Technology, 76131 Karlsruhe, Germany}
\author{John M. Martinis}
\affiliation{Google, Santa Barbara, California 93117, USA}
\author{Alexey V. Ustinov}
\affiliation{Physikalisches Institut, Karlsruhe Institute of Technology, 76131 Karlsruhe, Germany}
\affiliation{National University of Science and Technology MISI, Moscow 119049, Russia}
\date{\today}


\begin{abstract}
Superconducting integrated circuits have demonstrated a tremendous potential to realize integrated quantum computing processors.
However, the downside of the solid-state approach is that superconducting qubits  suffer strongly from energy dissipation and environmental fluctuations caused by atomic-scale defects in device materials.
Further progress towards upscaled quantum processors will require improvements in device fabrication techniques which need to be guided by novel analysis methods to understand and prevent mechanisms of defect formation.\\
Here, we present a technique to analyse individual defects in superconducting qubits by tuning them with applied electric fields. This provides a spectroscopy method to extract the defects' energy distribution, electric dipole moments, and coherence times. Moreover, it enables one to distinguish defects residing in Josephson junction tunnel barriers from those at circuit interfaces. We find that defects at circuit interfaces are responsible for about 60\% of the dielectric loss in the investigated transmon qubit sample. About 40\% of all detected defects are contained in the tunnel barriers of the large-area parasitic Josephson junctions that occur collaterally in shadow evaporation, and only $\approx$ 3\% are identified as strongly coupled defects which presumably reside in the small-area qubit tunnel junctions.
\\
The demonstrated technique provides a valuable tool to assess the decoherence sources related to circuit interfaces and to tunnel junctions that is readily applicable to standard qubit samples.

\end{abstract}

\maketitle 
\setlength{\parskip}{-0.25cm}

\section*{Introduction}
Superconducting qubits are implemented from resonant modes in non-linear electric microcircuits tailored from superconducting inductors, capacitors, and Josephson tunnel junctions~\cite{WendinReview,Devoret:Science:2013}.
Prototype quantum processors comprising a few tens of coupled qubits have already demonstrated quantum simulations of small molecules~\cite{IBM17}, error correction~\cite{QEC16}, and complex algorithms~\cite{Rigetti19}.
However, progress during the past decade was mostly achieved by improved circuit designs that reduced the interaction strength of qubits with material defects that are limiting device coherence~\cite{Wang:APL:2015,Gambetta17}. Since this strategy has seemingly exhausted its potential, further advancement will require an intense effort to understand and prevent the formation of defects in device fabrication.\\

Electric-field tuning of defects coupled to a superconducting resonator has been used to obtain information on their electric dipole moment sizes and densities~\cite{Sarabi16}, to control their population via Landau-Zener transitions~\cite{Khalil14}, and to realize microwave lasing~\cite{Rosen16}. Moreover, it served to investigate decoherence sources in disordered superconductors~\cite{leSuer18}, and AC-electric field modulation of defect baths may provide a path to decouple them from a resonator or qubit~\cite{Shlomi19}.\\

\begin{figure}[htb!]
	\includegraphics[width=9cm]{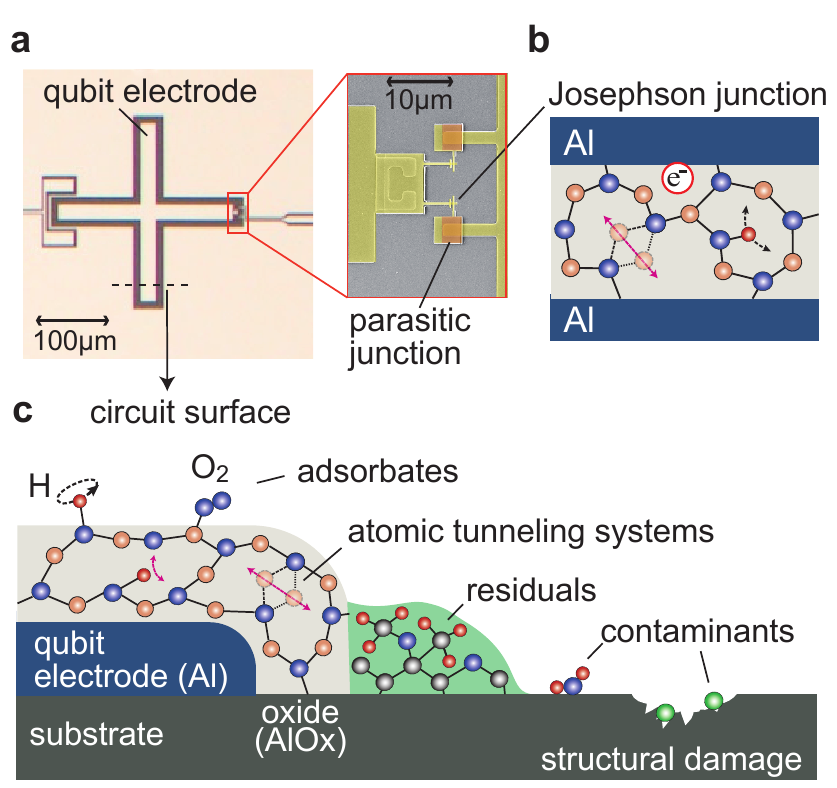}
	\caption{\textbf{Overview of defect types in superconducting qubits.} \textbf{a} Photograph of the cross-shaped capacitor electrode of a transmon qubit that is connected to the ground plane via Josephson junctions~\cite{Barends13}. Inset: the small junctions are contacted via large-area parasitic junctions (shaded orange).  \textbf{b} Illustration of defect types in the amorphous AlO$_x$ tunnel barrier of a Josephson junction, indicating atomic tunnelling systems, hydrogen impurities, and trapped electrons. \textbf{c} Sketch of surface defects, showing a cross-section of the qubit electrode and its native aluminum oxide which hosts structural TLS (not to scale). In addition, adsorbates such as hydrogen (H) and molecular oxygen (O$_2$) provide surface spins. Fabrication residuals such as photoresist, atmospheric contaminants, and substrate surface amorphization due to circuit patterning are further sources of surface defects.}
	\label{fig:models}
\end{figure}

There is a variety of microscopic models of how defects may emerge in quantum circuits, some of which are illustrated in Fig.~\ref{fig:models}.
A prominent class are structural tunnelling systems known from glasses, which are formed by a single or a few atoms tunnelling between two configurations in a disordered material~\cite{Phillips87,Anderson:PhilMag:1972}. This creates parasitic quantum two-level systems (TLS) which couple via their electric dipole moment to the oscillating electric fields in quantum circuits. Due to the material's random structure, TLS resonance frequencies are widely distributed, and those that are near resonance with qubits can dominate qubit energy relaxation~\cite{Martinis:PRL:2005}. 
Moreover, the thermal activation of low-energy TLS causes resonance frequency fluctuations of high-energy TLS, resonators, and qubits, which occur on time-scales spanning from milliseconds to hours and days~\cite{FaoroIoffe15,Mueller:2014, schlor2019,klimov2018,meissner2018,burnett2019}.
For quantum processors, this implies that each qubit needs to be frequently recalibrated, while individual qubits can also become completely unusable due to randomly occurring resonant interaction with fluctuating TLS. \\

Parasitic atomic tunnelling systems are contained in the amorphous aluminum oxide that grows natively on qubit circuit electrodes and which is used as a tunnel barrier in Josephson junctions. It is also assumed that structural tunnelling defects emerge from fabrication residuals such as photoresist~\cite{Quintana:APL:2014}, from impurity atoms such as interstitial hydrogen~\cite{Gordon:SciRep:2014,holder2013}, and due to damage of crystalline substrates by inadequate cleaning or film patterning procedures~\cite{dunsworth2017}. A further source of defects are surface adsorbates such as hydrogen atoms and O$_2$-molecules, whose unpaired spins have long been blamed as the major source of low-frequency (1/$f$-) noise in DC-SQUIDs and flux qubits~\cite{Anton:PRL:2013}, and which were recently reported to contribute also to charge noise~\cite{deGraaf:2017}.

\section*{Results}
\begin{figure*}
	\includegraphics[scale=1]{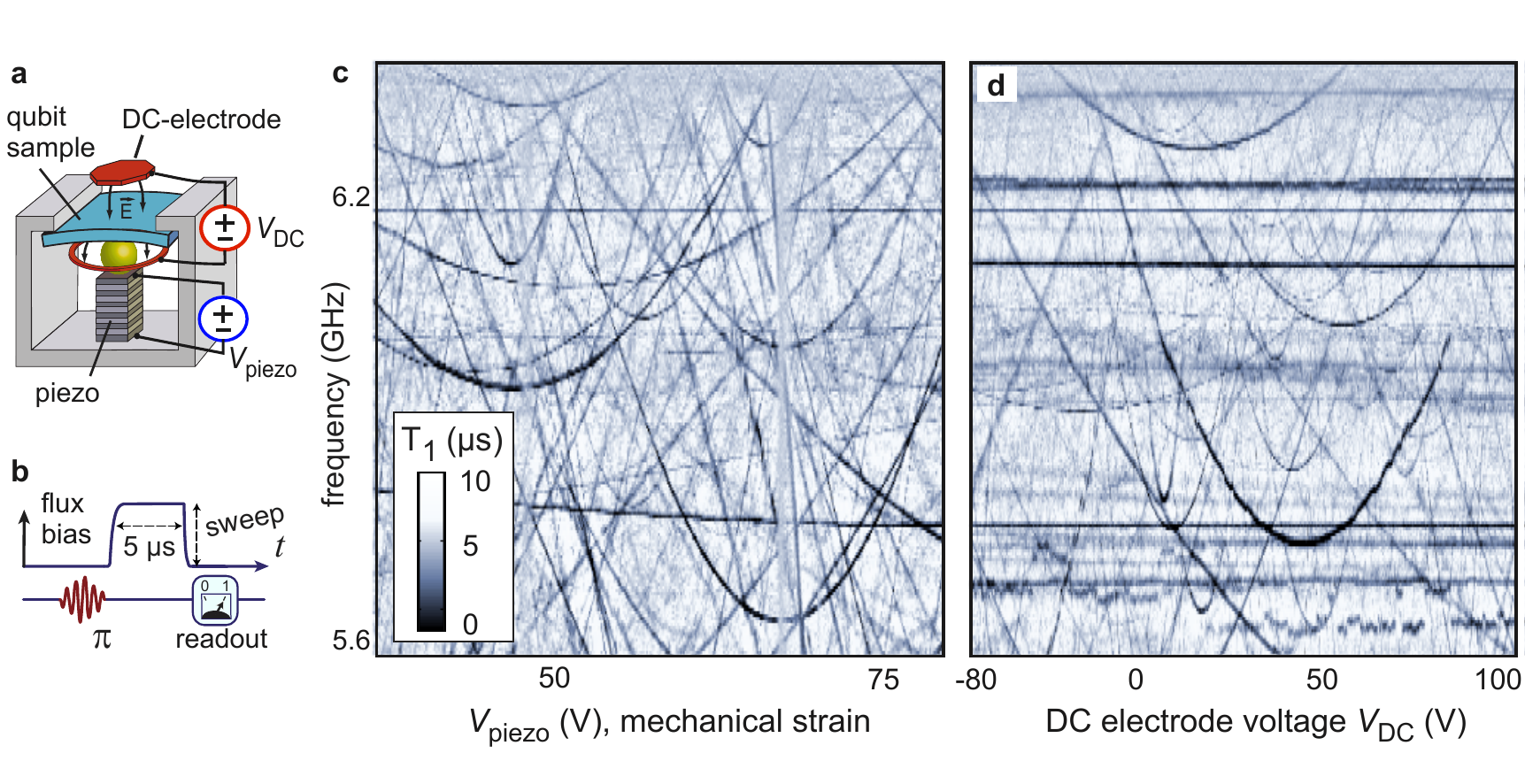}
	\caption{\textbf{Tuning defects by mechanical strain and electric field.}
		\textbf{a} Sketch of the setup for defect manipulation. The mechanical strain is controlled by the voltage $V_\mathrm{piezo}$ applied to a piezo actuator which slightly bends the qubit chip. The electric field is generated by two electrodes connected to a DC-voltage source $V_\mathrm{DC}$. \textbf{b} Pulse sequence realizing defect spectroscopy by measuring the frequency-dependence of the qubit's energy relaxation time $T_1$.
		\textbf{c} TLS resonance frequencies (dark traces indicating reduced $T_1$ time due to resonant interaction with defects) in dependence of applied mechanical strain. The horizontal line at 6.18 GHz is the resonance of a second qubit on the same chip.
		\textbf{d} Electric-field dependence of TLS resonance frequencies, plotted as a function of the voltage $V_\mathrm{DC}$ applied between the two electrodes. Hyperbolic traces stem from surface TLS at film interfaces, while horizontal lines reveal TLS residing in the tunnel barrier of a Josephson junction where they are screened from the electric field.}
	\label{fig:strainfield}
\end{figure*}

In this article, we describe a technique to manipulate surface defects which are residing at qubit circuit interfaces such as the substrate-metal, the substrate-vacuum, or the metal-vacuum interfaces. For this, the sample is exposed to a global DC-electric field generated by electrodes surrounding the chip as illustrated in Fig.~\ref{fig:strainfield}\textbf{a}. A surface defect responds to the field with a change of its asymmetry energy $\varepsilon$, which together with its (constant) tunnelling energy $\mathit{\Delta}$ determines its resonance frequency, $f_\mathrm{TLS} = \sqrt{\mathit{\Delta^2} + \varepsilon^2} / h$. 
The global DC-electric field does not generate any field in the junction's tunnel barriers, because the small charging energy of the transmon qubit allows Cooper-pairs to compensate any induced charge by tunnelling off or onto the island~\cite{Schreier08}. Thus, TLS inside the tunnel barrier of a small or large (parasitic) junction do not respond to the applied external electric field and can hereby be distinguished from surface defects.\\
In addition, our setup allows us to  tune all defects, irrespective of their location in the circuit, by mechanical strain that is generated via a piezo actuator slightly bending the qubit chip~\cite{Grabovskij12}. In earlier experiments on phase qubits, the strain-tuning technique has proven useful to reveal mutual TLS interactions~\cite{Lisenfeld2015} and to probe the coherent quantum dynamics of TLS to quantify their coupling to the bath of thermally fluctuating defects~\cite{Lisenfeld2016,meissner2018}. Here, we use it to verify that also defects that do not respond to the electric field behave according to the standard tunnelling model, and to enhance the number of observable defects that are not tuned by the electric field.\\

In our experiments, we detect defects by measuring the frequency-dependence of the qubit's energy relaxation rate $1/T_1$, where local maxima indicate enhanced dissipation due to resonant interaction with individual TLS. A fast method uses the swap-spectroscopy protocol shown in Fig.~\ref{fig:strainfield}\textbf{b}, where the qubit is populated by a microwave $\pi-$pulse, tuned to various probe frequencies using a variable-amplitude flux pulse, and afterwards read out~\cite{Cooper04,Lisenfeld2015}. To economize time, we use a fixed flux pulse duration and take only two further reference measurements at zero flux pulse amplitude to estimate the energy relaxation rate (see supplementary material for details). The flux pulse duration is set to about half the qubit's T$_1$ time in order to balance signal loss against sensitivity to weakly interacting defects.\\

Figure~\ref{fig:strainfield} presents exemplary data taken with a transmon qubit in Xmon-geometry that was fabricated at UCSB as described in Ref.~\cite{Barends13}. The strain-dependence of the qubit's $T_1$ time, plotted as a function of the applied piezo voltage $V_\mathrm{piezo}$ in Fig.~\ref{fig:strainfield}\textbf{c}, shows that the resonance frequency of all detectable TLS follows the expected hyperbolic dependence. In contrast, the horizontal lines in the electric field dependence Fig.~\ref{fig:strainfield}\textbf{d} reveal a subset of defects that don't respond to the applied field as it is expected if they reside inside the tunnel barrier of a Josephson junction. Simulations of the induced electric fields (see supplementary material) were used to verify that all detectable surface defects, i.e. those that couple sufficiently strongly to the field induced by the qubit's plasma oscillation, should also respond to the applied field. \\

\begin{figure*}
	\includegraphics[scale=1]{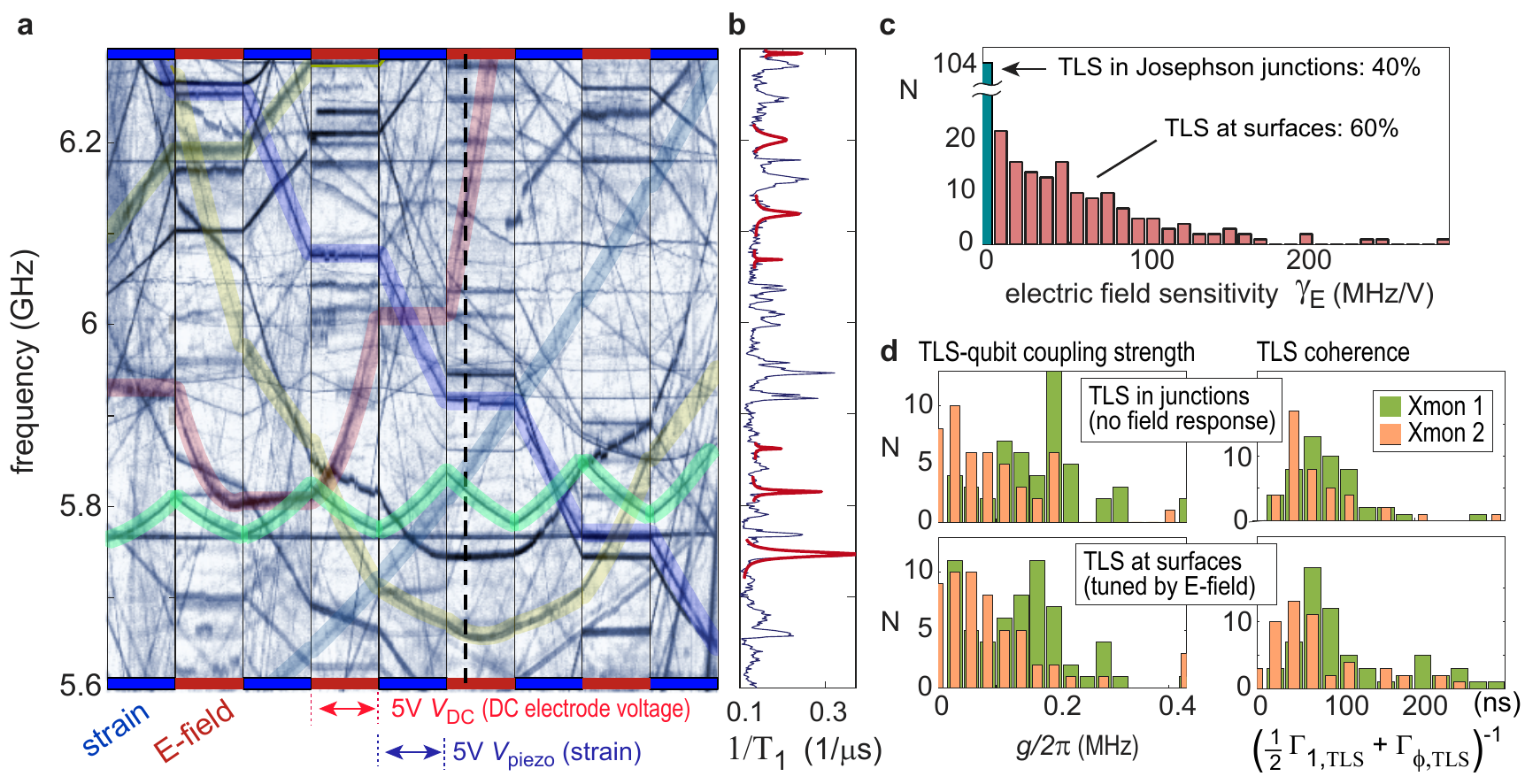}
	\caption{\textbf{Statistics on the response to strain and electric field.} \textbf{a} Alternating measurements of the mechanical strain (blue margins) and electric field dependence (red margins) of TLS resonance frequencies. Coloured trace highlights follow exemplary fits from which the TLS' deformation potential $\gamma_S$ and field coupling strength $\gamma_E$ are obtained. \textbf{b} Vertical cross-section of the data shown in \textbf{a} (black dashed line), recalculated to the energy relaxation rate $1/T_1$. Lorentzian fits to individiual TLS resonances Eq.~(\ref{eq:tlsresonance}) result in qubit-TLS coupling strengths $g$ and TLS decoherence rates. The frequency step was 1 MHz. \textbf{c} Distribution of TLS sensitivities to electric field $\gamma_E$. No response ($\gamma_E <$ 1 kHz/V) is observed in 104/260 TLS (40$\%$.).
		\textbf{d} Histograms of TLS-qubit coupling strengths $g$ (left column) and TLS coherence times $(\frac{1}{2} \Gamma_\mathrm{1, TLS} + \Gamma_\mathrm{\Phi, TLS})^{-1}$ (right column), plotted separately for TLS that do not respond to the field (top row) and for field-tunable TLS (bottom row). Similar coherence times of 50 - 100 ns are observed irrespective of the TLS' response to electric field. The relatively small coupling strengths up to $g \approx 0.2~ $MHz indicate that nearly all defects which do not respond to the E-field are residing in the large-area parasitic junctions rather than the small qubit junctions.
	} 
	\label{fig:multispec}
\end{figure*}

In order to characterize the TLS' response to the two types of stimuli more systematically, we alternate between strain and electric field sweeps. This results in data as shown in Fig.~\ref{fig:multispec}\textbf{a}, where the strain was linearly increased in segments with blue margins and the E-field was swept in red-framed segments. To extract the TLS' sensitivities to strain $\gamma_S$ and to electric field $\gamma_E$, we fit each visible trace to the hyperbolic resonance frequency dependence with the TLS' asymmetry energy $\varepsilon(V_\mathrm{piezo}, V_\mathrm{DC}) = \varepsilon_i + \gamma_S \cdot V_\mathrm{piezo} + \gamma_E \cdot V_\mathrm{DC}$. Here, $\gamma_E$ is proportional to the TLS' electric dipole moment component that is parallel to the local electric field, and $\varepsilon_i$ is an intrinsic offset energy. A few exemplary fits are indicated by highlighted traces in Fig.~\ref{fig:multispec}\textbf{a}. The distribution of extracted field sensitivities $\gamma_E$ is plotted in Fig.~\ref{fig:multispec}\textbf{c}. We find that from a total of 260 observed TLS in two qubits, 40\% do not respond to the electric field and thus reside in tunnel barriers - most likely in the large-area parasitic junctions as explained below. In contrast, all observed defects are tuned by mechanical strain as expected from the standard tunnelling model.\\

The lineshape of the TLS resonances observed in $T_1$-spectroscopy contains further information about the TLS' decoherence rate and dipole moment size. After recalculating the data in Fig.~\ref{fig:multispec}\textbf{a} into the qubit energy relaxation rate $1/T_\mathrm{1}$, each Lorentzian resonance is fitted to the equation~\cite{Barends13}
\begin{equation}
1/T_\mathrm{1} = \frac{2\ g^2 \Gamma}{\Gamma^2 + \delta^2} + \Gamma_\mathrm{1,Q},
\label{eq:tlsresonance}
\end{equation}
where $\Gamma = (\Gamma_\mathrm{1,TLS}/2 + \Gamma_\mathrm{\Phi,TLS}) + (\Gamma_\mathrm{1,Q}/2 + \Gamma_\mathrm{\Phi,Q})$ is the sum of TLS and qubit energy relaxation and dephasing rates, and $\delta$ is their detuning. The coupling strength $g' = (\vec{p}\, \cdot \vec{E})/\hbar$ between the qubit and a defect is the scalar product of the defect's electric dipole moment $\vec{p}$ and the local electric field $\vec{E}$ induced by the qubit's plasma oscillation. Figure~\ref{fig:multispec}\textbf{b} shows exemplary fits to Eq.~(\ref{eq:tlsresonance}) along the data marked by the black dashed line in Fig.~\ref{fig:multispec}\textbf{a}. We note that this provides the effective coupling strength $g = g' \cdot \mathit{\Delta}/\sqrt{\mathit{\Delta}^2+\varepsilon^2}$ that is dressed by the TLS' matrix element which is typically unknown, but can be measured with strain- or E-field-spectroscopy if the TLS' tunnel energy $\mathit{\Delta}$ lies within the qubit's tunability range.\\

\begin{figure}
	\includegraphics[scale=1]{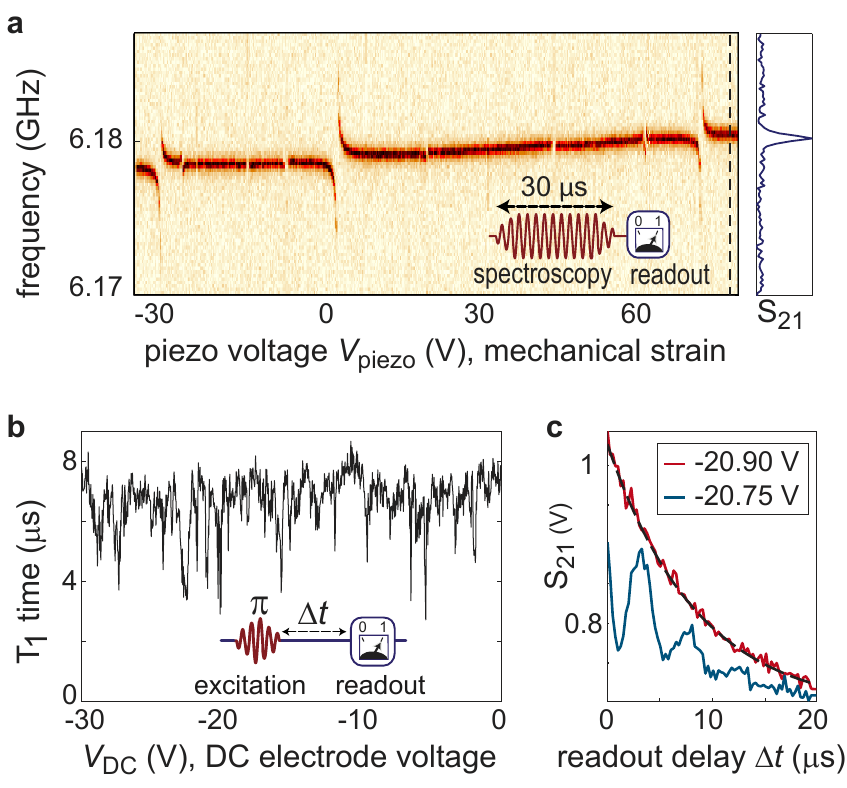}
	\caption{\textbf{Detection of strongly coupled defects.} \textbf{a} Qubit spectroscopy, taken by a long microwave pulse of varying frequency (vertical axis) to observe the qubit resonance (red color, see also right inset). Variation of the mechanical strain (horizontal axis) reveals avoided level crossings due to strongly coupled TLS that are tuned through the qubit resonance. \textbf{b} Qubit $T_1$-time measurement as a function of the applied DC-electric field. Lorentzian dips indicate resonant coupling to surface TLS. \textbf{c} In about 3$\%$ of observed dips in the qubit's $T_1$ time, damped oscillations are observed (blue curve) which herald quantum state swapping between the qubit and coherent surface TLS. In all other cases, purely exponential decay is found (red curve, fitted by black dashed line).
	} 
	\label{fig:cw}
\end{figure}

In Fig.~\ref{fig:multispec}\textbf{d}, the distributions of TLS-qubit coupling strengths and TLS coherence times are plotted separately for junction-TLS that do not respond to the electric field (top row) and for field-tunable TLS residing at circuit interfaces (bottom row). Both classes show very similar qubit-TLS coupling strengths below $g/2\pi\approx$ 0.4 MHz and typical TLS coherence times of 50 - 100 ns, in agreement with independent measurements on identically fabricated samples~\cite{Barends13}. On average, surface TLS are slightly stronger coupled to the qubit Xmon~1 because of its smaller gap between the qubit island and the ground plane as compared to Xmon~2 (see supplementary material).
\\
\section*{Discussion}
Considering field-tunable defects,  the measured coupling strengths around $g/2\pi\approx$ 0.1 MHz correspond to parallel dipole moment components of 0.2 - 0.4 e\AA\ when they reside in the vicinity of the qubit electrode's edge, where simulations indicate a plasma oscillation field strength of $|E|$=10 - 20 V/m (see supplementary material).  
Measurements in AlO$_x$~\cite{Martinis:PRL:2005,Muller:2017} and SiN$_x$~\cite{Sarabi16} have found similar dipole moment sizes.
\\
If the defects that do not respond to the applied E-field were inside the tunnel barrier of the qubit's small Josephson junctions and had dipole moments of similar magnitude, one would expect much larger coupling strengths up to $g/2\pi\approx40$ MHz in contrast to our observation Fig.~\ref{fig:multispec}\textbf{d}. However, in our sample, each qubit junction is connected to the ground plane via an additional larger "parasitic" or "stray" junction as a consequence of the employed shadow evaporation technique, see Fig.~\ref{fig:models}\textbf{a}. Since these junctions are connected in series, the voltage drop across each is inversely proportional to its area, resulting in a factor of $\approx$ 200 smaller electric fields in the larger parasitic junctions. TLS in these junctions thus couple at a strength of only $g/2\pi\approx0.1$ MHz when they have dipole moments of 0.3 e\AA\ which is in excellent agreement with our measurements presented in Fig.~\ref{fig:multispec}\textbf{d}.
Considering that junction and surface defects show similar coupling strengths and coherence times, we can conclude that in our sample $40\%$ of the qubit's dielectric loss is due to defects in the parasitic Josephson junctions. This clearly shows the importance to avoid parasitic junctions by advanced fabrication methods such as the 'bandaging' technique where they are shorted by an additional metallization layer~\cite{dunsworth2017}.
\\

The number of observable TLS resonances, averaged over all applied values of E-field and strain, was $N_p$=15/GHz for TLS in the parasitic junctions and $N_s$=23/GHz for surface TLS. For the parasitic junctions, this corresponds to a TLS density of $P_{0,p}=250\, \mathrm{GHz}^{-1}\mu m^{-3}$ when assuming a tunnel barrier thickness of 2.5 nm. This density agrees with values in the range of $200 - 1200 \,\mathrm{GHz}^{-1}\mu m^{-3}$ typically quoted for bulk dielectrics~\cite{FaoroIoffe2015}, and corresponds to a dielectric loss tangent~\cite{Gao08} $\tan \delta_{0,p} = \pi P_{0,p}\, p^2 (3 \epsilon_0 \epsilon_r)^{-1} \approx 1.8\cdot 10^{-4}$, where we chose $p = 0.4 e\mathrm{\AA}$ and $\epsilon_r\approx10$ for AlO$_x$. This value is a factor of $\approx$ 10 smaller than typically quoted~\cite{Martinis:PRL:2005,Deng14}, presumably because it is based on the number of TLS traces that are clearly visible in swap spectroscopy, while contributions from weakly coupled TLS are not taken into account.\\
If the detected surface defects were distributed uniformly along the $\approx1.4$mm-long edge of the qubit island, their average distance would be 60 $\mu$m. However, there are reasons to assume that surface defects may accumulate in the vicinity of the Josephson junctions where additional lithographic processing is required~\cite{dunsworth2017}. This can be clarified in future experiments by employing a laterally positioned DC-electrode.\\

Defects in the small tunnel junctions can be identified by their strong coupling to the qubit. This gives rise to avoided level crossings, which we reveal by monitoring the qubit's resonance peak as a function of the applied strain. Figure~\ref{fig:cw}\textbf{a} shows exemplary data which typically display up to 3 splittings larger than 5 MHz in our accessible strain range.
Comparing this number to the $\approx 40$ TLS per qubit found in the parasitic junctions, the result comes closer to the ratio of the junction's circumferences ($\approx 14$) than their area ratio ($\approx 200)$. This gives a hint that defects might predominantly emerge at junction edges.\\

Finally, we test the coherence and coupling strength of surface TLS directly by observing coherent swap oscillations in resonantly coupled qubit-TLS systems~\cite{Cooper04}. For this, the qubit population decay is observed in a $T_1$-time measurement for a range of applied electric fields. Figure~\ref{fig:cw}\textbf{b} shows the qubit's $T_1$ time extracted from fits to exponential decay curves, which displays minima whenever the applied field tuned a TLS into resonance with the qubit. In about 3$\%$ of more than 100 investigated $T_1$-minima, we observed coherent swap oscillations with frequencies $g/2\pi$ = 150 - 520 kHz as shown in Fig.~\ref{fig:cw}\textbf{c}. This small subset of surface TLS thus has coherence times of a few microseconds, probably because they were close to their symmetry points where dephasing is suppressed~\cite{Lisenfeld2016}. \\

To summarize, our results confirm that defects residing at the substrate-metal, substrate-vacuum, and metal-air interfaces are a limiting factor of coherence in contemporary superconducting qubits. Parasitic (stray) junctions must be avoided since they host large numbers of defects. So far, all experiments studying individual defects in superconducting qubits can be explained by a single type of defect having typical electric dipole moments $|\vec{p}| \approx$ 0.1 - 1 e\AA \ and coherence times of typically 100 ns, in seldom cases extending up to a few $\mu$s.\\
The presented technique distinguishes defects in the tunnel barriers of Josephson junctions from those at surfaces. It provides a diagnostic tool to improve device fabrication by assessing the quality of tunnel junctions and circuit surfaces. This method can be easily implemented and applied to a variety of superconducting qubit types. Actually, a single electrode that is biased against the on-chip ground plane is sufficient to clearly distinguish surface-TLS from tunnel barrier defects.\\

Using multiple separately biased electrodes, it becomes possible to pinpoint the location of individual defects by comparing their response to simulations of the spatially dependent electric fields. As we will describe elsewhere, our current setup using two electrodes as sketched in Fig.~\ref{fig:strainfield}\textbf{a} already suffices to distinguish defects on the substrate-metal, the substrate-vacuum, and the metal-vacuum interfaces. Lastly, we note that integration of on-chip electrodes or gates for direct voltage-biasing of transmon qubit islands offers a way to \emph{in-situ} mitigate decoherence in qubits that happen to be near resonance with a surface defe,ct, facilitating the path towards scaled-up quantum processors.

\subsection*{Further Details}
The supplementary material contains further details on spectroscopy methods, E-field  simulations, and plots of complete data sets from strain- and E-field-dependent defect spectroscopy.

\subsection*{Acknowledgements}
JL is grateful for funding from the Deutsche Forschungsgemeinschaft (DFG), grant LI2446-1/2. AB acknowledges support from the Helmholtz International Research School for Teratronics (HIRST) and the Landesgraduiertenförderung-Karlsruhe (LGF). AVU acknowledges support by the Ministry of Education and Science of the Russian Federation NUST MISIS (contract No. K2-2017-081). This work was supported by Google.

\subsection*{Author contributions}
JL performed the measurements and data analysis and wrote the manuscript. AB implemented the setup for defect tuning by mechanical strain and electric field, and performed simulations. The qubit sample was developed and fabricated by partners from Google. All authors discussed the results and commented on the manuscript.

\clearpage

\section{Supplementary Methods}
\subsection{Xmon qubit sample}
The sample measured in this work was fabricated by Barends et al. as described in Ref.~\cite{Barends13}. We measured a chip containing three uncoupled Xmon qubits which differ by the geometry of the coplanar qubit capacitor, having a width of S=16 (24) $\mu$m  for sample "Xmon1" ("Xmon2"), while the gap to the ground plane had a width of 8 (12) $\mu$m, respectively.\\
The qubit island,  groundplane, and resonators are patterned from aluminum deposited by molecular beam epitaxy on a sapphire substrate. The Josephson junctions of the DC-SQUID connecting the qubit island to ground are fabricated by eBeam shadow evaporation. Each arm of the DC-SQUID contains a small tunnel junction of size 0.3 x 0.2 $\mu m^2$ in a series connection with a larger "parasitic" junction of size 4.0 x 2.9 $\mu m^2$. The function of the parasitic junctions is to connect the small junctions to the ground plane, and their large area ensures negligible contributions to the qubit's charging energy and Josephson phase dynamics. The total capacitance of all investigated qubits (estimated from a measurement of their anharmonicity) is $\approx$ 86 fF. Further details are contained in~\cite{Barends13} and its supplementary material.\\

Several years have passed between sample fabrication and our measurements, during which the sample was stored at room temperature in a nitrogen-enriched atmosphere while it was covered with photoresist (Microposit SPR955). Sample aging due to incorporation of contaminants from the atmosphere and the photoresist may explain why we observed shorter $T_1$ times (on average about 8 - 9 $\mu$s for both qubits) compared to the earlier measurements in a different setup~\cite{Barends13}. Moreover, since the sample was cleaned only by an acetone-wash prior to our measurements, photoresist residuals may add to a reduction of coherence times and enhance the number of surface TLS observed in our sample. For these reasons, the observed defect densities are not representative of the quality of the fabrication process nor do they allow us to draw conclusions on the effect of sample aging.
While a reduced qubit coherence time does not affect our conclusions on the relative distributions of defects among circuit interfaces and the tunnel barriers of small- and  large-area parasitic junctions, it does increase the minimum coupling strength of defects that can be detected in $T_1$-time spectroscopy.

\subsection{Experimental setup}
The sample was measured in an Oxford Kelvinox 100 wet dilution refrigerator at a temperature of 30 mK. The qubit chip was installed in a light-tight aluminum housing surrounded by a cryoperm magnetic shield. To avoid stray infrared (IR) radiation leaking into the sample chamber via the Teflon insulation of SMA connectors and coaxial cables, we installed custom-made IR filters consisting of a short ($\approx$ 2cm) section of coaxial cable in which the Teflon dielectric was replaced by black epoxy (Stycast 2850 FT). However, due to the space constraints in the available cryostat, we did not employ a multi-layer "box-in-a-box" shielding approach~[S1] to further minimize stray IR radiation, which might be a further reason why our measured qubit coherence times were smaller compared to the earlier measurements in a different setup by Barends et al.~\cite{Barends13}.\\

A standard homodyne microwave detection setup was used to read out the qubit state by measuring the dispersive shift of a readout resonator capacitively coupled to the qubit and a common transmission line. Figure~\ref{fig:setup} illustrates the employed setup.  In this work, we deduced the qubit population from the measured transmission amplitude $|S_{21}| = \sqrt{V_I^2+V_Q^2}$ near the resonance frequency of the readout resonator, where $V_I$ and $V_Q$ are the voltages at the IF-ports of the downconverting IQ-mixer.\\
The DC-bias voltage applied to the two electrodes for electric-field tuning of defects was generated by custom-made electronics using an Arduino microprocessor as a computer USB-interface that generated a digital signal that was sent via an optical fiber (to avoid ground loops) to control a DAC714 16-Bit digital-to-analog converter circuit, whose output voltage was amplified by a factor of 50 using an operational amplifier. At the 4K-stage of the cryostat, this signal was strongly filtered using an RC-lowpass that had a cutoff frequency of about 10 kHz. A twisted pair of enameled wire was used to connect the DC electrodes.\\

Figure~\ref{fig:sampleholder}\textbf{a} shows the CAD-drawing of the sample housing and piezo holder. 
The DC-electric field for tuning of surface defects is generated by electrodes above and below the chip as shown in Figs.~\ref{fig:sampleholder}\textbf{b} and \textbf{c}. The top electrode consists of a copper-foil/Kapton foil stack that is glued to the lid of the sample box. The bottom electrode is patterned into the backside of the printed circuit board carrying the qubit chip, and has the form of a ring to leave space for the piezo exerting force to the centre of the qubit chip. Due to the chosen geometry of the DC electrodes, the electric field is not homogeneous on the scale of the chip's size. However,  finite element simulations~[S3] showed that E-field variations along the qubit island's film edge were below the simulation's precision for the qubit located in the centre of the chip (Xmon 1). In this work, E-field inhomogeneities would only affect the measurement of TLS' coupling strengths to the field taken on XMon 2 (Fig. 3c). This issue can be avoided by using larger DC electrodes.

\subsection{Simulations of electric fields}
The electric fields generated by the qubit plasma oscillation and the fields induced by the global electrodes are simulated using the finite element solver ANSYS Maxwell 2015 (release 16.2.0). Figures~\ref{fig:simuqubit}~\textbf{a} and \textbf{b} show the dependence of the qubit-induced electric field magnitude $|E_q|$ at the substrate-metal (SM) and substrate-vacuum (SV) interfaces, respectively. The spatial axes are defined in the insets. In agreement with Ref.~\cite{Barends13}, we find the field to decay according to a power law $|E_q| = a\cdot x^\beta$, where $a=25, \beta=-0.5$ and $a=50, \beta=-0.6$ for the SM and SV interfaces, respectively, as shown by the fits. The insets show the 2-dimensional dependence of the field strength, reaching values between 10 and 25 V/m at the edge of the qubit electrode. In Fig.~\ref{fig:simuqubit}\textbf{c} the field strength is plotted along the rounded edge of the qubit electrode at the oxide-vacuum interface (OxV) and inside the native oxide layer (Ox).\\
The simulation also shows that only TLS residing within a distance of $\approx 200$nm to the edge of the qubit island experience field strengths that couple them strongly enough to the qubit to be detected with our sample. For example, a TLS having an (unusually) large electric dipole moment of 10 D and residing at $x >$ 200nm would couple it to the qubit by $g/2\pi <$ 0.05 MHz, which is about the typical detection sensitivity of the UCSB samples given their coherence times. Thus, 200 nm is the maximum distance to the edge where such large TLS are still detectable.\\

The electric field induced near the qubit island's film edge is shown in Fig.~\ref{fig:Efield_tb} for the case when a voltage of 1V is applied between the top and bottom electrodes. This also shows that in the substrate and the (4nm-thick) AlOx on the electrode, the field is reduced by about a factor of 10 due to the material's permittivity. The maximum field strength occurs at the common edge between the substrate, film, and vacuum, and reaches values up to 3 kV/m in this case.\\
The finite element simulations of the electric field also confirmed that all TLS that are detectable by the qubit should also be tuned by the field induced by the global electrodes. This means that within the distance from the qubit edge where TLS are detectable, the absolute strength of the globally applied E-field as well as its projection onto the TLS electric dipole moment is large enough to result in detectable detuning.

\subsection{Defect spectroscopy}
The resonance frequencies of defects are detected by measurements of the frequency-dependent qubit energy relaxation rate which displays Lorentzian peaks due to dissipation from resonant defect interaction~\cite{Cooper04,Lisenfeld2015}. To economize time, we do not track the full exponential decay curve but rather calculate the $T_1$ time using the sequences depicted in the insets of Fig.~\ref{fig:swapspec}\textbf{a}. Sequence \textbf{b} is used to calibrate the maximum signal $P_1$ for the excited qubit, and the background signal for the qubit in the ground state $P_0$ is obtained by omitting the qubit excitation pulse as shown in Fig.~\ref{fig:swapspec}\textbf{c}. The signal $P_s$ (see Fig.~\ref{fig:swapspec}\textbf{d}) gives the remaining qubit population after it had time to interact with defects at different frequencies selected by the flux pulse amplitude. The qubit $T_1$ time is then calculated from the exponential decay law,
\begin{equation*}
T_1 = T_s / \log \left ( \frac{P_1 -P_0}{P_s - P_0} \right ),
\end{equation*}
where $T_s$ is the swap pulse duration that is chosen as $T_s \approx 0.5\, T_1$ to balance signal loss against sensitivity to detect also weakly interacting defects.\\
To estimate the uncertainty of the fast 3-point method to determine $T_1$, we simulated decay curves with qubit and TLS parameters similar to typical values and of comparable signal-to-noise ratios as in the experiment. Figure~\ref{fig:swapspec}\textbf{e} compares the distribution of $T_1$ times obtained using a full exponential fit including 30 data points (red) with the results of the 3-point method (blue). The distributions of parameters estimated from Lorentzian dips due to the resonance with a TLS of nominal coherence time of 100 ns and coupling strength $g=100$ kHz are shown in Fig.~\ref{fig:swapspec}\textbf{f} and \textbf{g}. The corresponding standard deviations $\sigma$ are quoted in the legends and are approximately twice larger for the 3-point method as for full exponential fits.\\

We operate the pulse generator (Tektronix AWG 5914B) in the sequence-mode, where all sequences shown in the insets of Fig.~\ref{fig:swapspec}\textbf{a} are interleaved. We use a frequency resolution of typically 1 MHz, which results in a number of 700 different sequences for the data shown in Fig.~\ref{fig:Xmon2_supp}, and a repetition rate of 25 kHz (limited by the qubit's $T_1$ time) at typically 1000 averages, resulting in about 30 seconds required for each complete $T_1$ vs. frequency measurement. A full data set spans a piezo voltage range of -30V to +100V (at a resolution of 0.1V) and a DC-electrode voltage range of -65 to 65 V (at a resolution of 0.1 V). In addition, the qubit resonance frequency and resonator readout frequency are recalibrated at regular intervals (20 minutes) by an automatic procedure to compensate small drifts. The total time required to obtain a dataset as shown in Fig.~\ref{fig:Xmon2_supp} is about 33 hours.\\

Precise calibration of the flux pulse for qubit detuning is vital to obtain sharp resonances and high contrast in $T_1$ spectroscopy. Typically, impedance mismatches at cold attenuators, plugs, or filters in the coax cables result in signal overshoot and ringing. To correct for this, we use a similar technique as described in the Supplementary Material to Ref.~\cite{Barends13}, where the shape of the flux pulse reaching the qubit is calibrated $\emph{in-situ}$ by shifting a nominally rectangular flux pulse through a short microwave pulse that excites the qubit. The measured resonance frequency of the qubit as a function of delay between flux and excitation pulses is then used to deduce the time-dependence of the flux pulse. Deviations from the intended rectangular pulse shape are extracted as an error signal which is then subtracted from the pulse generated by the arbitrary waveform generator. If necessary, we use a second pass to further reduce small ripple. The effect of inadequate pulse calibration can be noticed in Fig.~\ref{fig:Xmon1_supp}\textbf{a} below a frequency of 6 GHz, resulting in increased widths of TLS traces.\\

The qubit-TLS coupling strengths and TLS coherence times are obtained by fitting the Lorentzian dips in $T_1$ spectroscopy as described in the main text (Fig.~3\textbf{b}). To reduce fit uncertainties, resonances for each visible TLS trace were fitted at several (typically 5) different combinations of applied strain and electric field, and then averaged. These fits were preferentially performed on Lorentzian dips near the resonance frequency of the unbiased qubit in order to minimize effects of improper flux pulse calibration.\\

The measured resonance curve of TLS may be broadened due to their interactions with thermally activated defects at low frequencies whose switching time is smaller than the $\approx$ 30 seconds required to obtain one curve of qubit $T_1$ vs. frequency. Such fast thermal fluctuators can be characterized more precisely eg. by dedicated spin-echo measurements~\cite{Lisenfeld2016} and by using high-frequency TLS as probes of their switching dynamics~[S4]. Several cases of TLS interactions with slower thermal fluctuators are seen eg. in Fig 3a, mostly as telegraphic switching of their resonance frequencies as expected when a single thermally activated TLS is involved, or as singular jumps if a metastable TLS was involved.	
In our measurements, the frequency resolution was about 1 MHz (i.e. the frequency step in the swap spectroscopy measurement).
\\


\section{Supplementary Figures}


Full data sets acquired in typical measurements of the TLS' response to mechanical strain and electric field are presented in Fig.~\ref{fig:Xmon2_supp} (for Xmon 2) and Fig.~\ref{fig:Xmon1_supp} (for Xmon 1).\\

Figure~\ref{fig:allstats} provides additional scatter plots of extracted TLS parameters. A correlation between the plotted value pairs can only be detected between $\gamma_E$ and $\gamma_S$, for which a Pearson correlation coefficient of about 0.4 is found (considering only TLS that do respond to the E-field). Such a correlation between the strain-coupling strength and the electric dipole moment of TLS is typically found for intrinsic defects in glasses, i.e. those originating in structural tunneling systems rather than due to incorporation of extrinsic (alien) atom species~[S5].  

\section{Supplementary Notes}
The here discussed work was part of the PhD studies of Alexander Bilmes at Karlsruhe Institute of Technology (KIT). Further details on the experimental setup, electric field simulations, and data acquisition can be found in his thesis~[S3].


\section{Supplementary References}
\noindent
[S1] Barends, R. et al., \textit{Minimizing quasiparticle generation
from stray infrared light in superconducting quantum circuits}, Applied Physics Letters \textbf{99}, 113507 (2011)\\

\noindent [S2] Barends, R. et al., \textit{Minimizing quasiparticle generation
from stray infrared light in superconducting quantum circuits}, Applied Physics Letters \textbf{99}, 113507 (2011)\\

\noindent [S3] Bilmes, A. \textit{Resolving locations of defects in superconducting transmon qubits}, Phd thesis, Karlsruhe Institute of
Technology (KIT) (2019), https://publikationen.bibliothek.kit.edu/1000097557\\

\noindent [S4] Meißner, S. M., Seiler, A., Lisenfeld, J., Ustinov, A. V., and
Weiss, G. \textit{Probing individual tunneling fluctuators with
coherently controlled tunneling systems}, Phys. Rev. B \textbf{97}, 180505 (2018)\\

\noindent [S5] Golding, B., Schickfus, M. v., Hunklinger, S. and Dransfeld,
K, \textit{Intrinsic electric dipole moment of tunneling systems
in silica glasses}, Phys. Rev. Lett. \textbf{43}, 1817–1821 (1979)

\renewcommand{\thefigure}{\textbf{S\arabic{figure}}}
\setcounter{figure}{0}

\begin{figure*}
	\includegraphics[width=12cm,height=13cm]{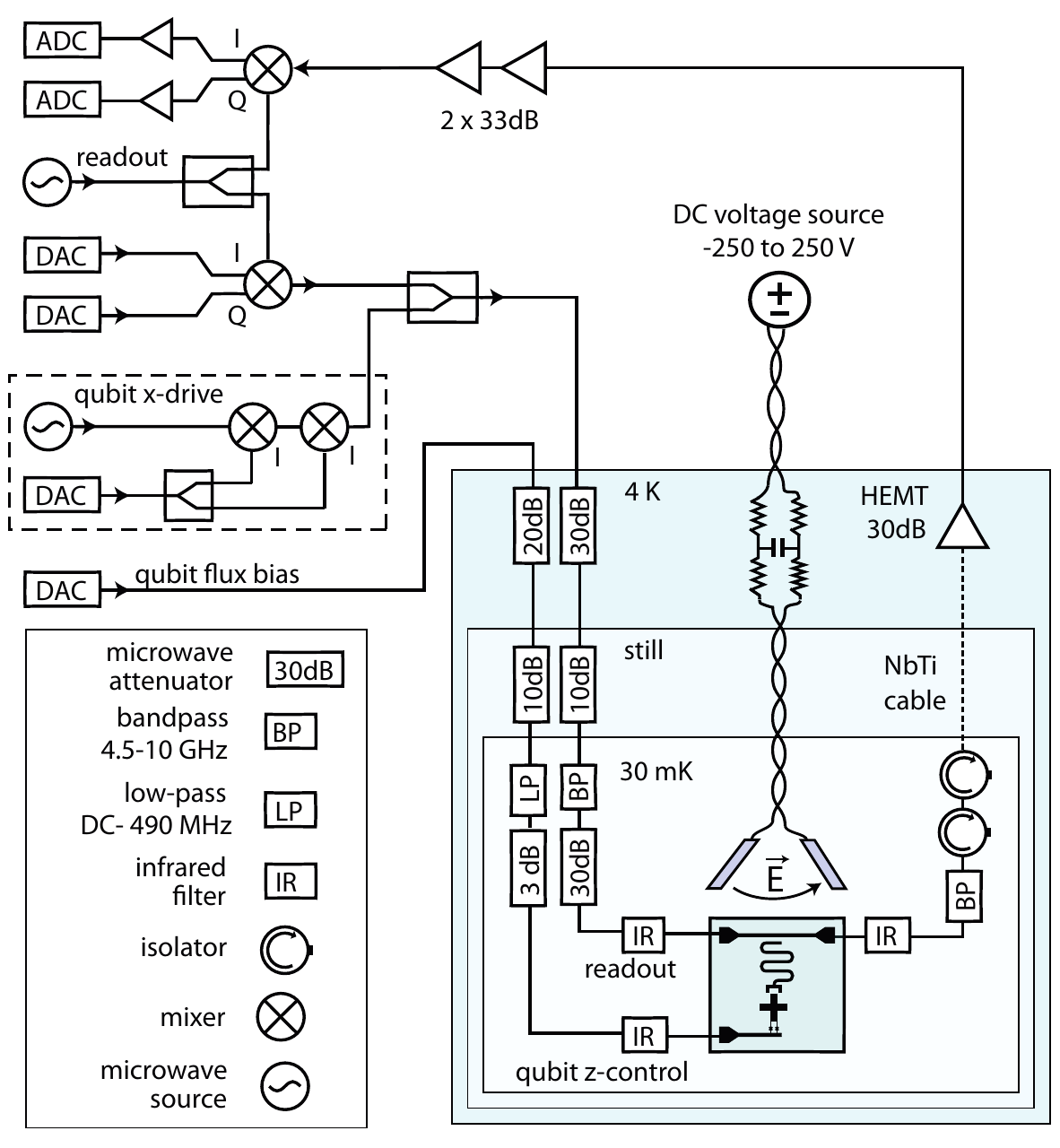}
	\caption{Setup schematic. Phase- and amplitude-controlled microwave pulses for qubit readout are generated by mixing the output of a continuous-wave microwave source (Agilent E8251A) with DC pulses from an arbitrary wave generator (DAC) (Tektronix AWG 5914B, 1.2 GS/s) using an IQ-Mixer (Marki IQ-4509LXP). Microwave pulses for qubit driving are shaped using two mixers (Marki M8-0420LS) in series to reduce leakage. 
	Thermalization of coax cables and noise protection is provided by attenuators thermally anchored to different temperature stages, band-pass (BP) and low-pass filters (LP) (Mini-Circuits), and custom-made infrared filters (IR). After interacting with the sample, the signal passes through two magnetically shielded isolators (QuinStar QCY-060400CM00), a high-mobility electron transistor (HEMT) cryogenic microwave amplifier (Low-Noise Factory LNF-LNC4-16A) and two room-temperature amplifiers (Mini-Circuits VA-183-S). It is then down-converted to DC by mixing it with a copy of the reference signal, amplified, and digitized using a fast sampling card (Spectrum M2i.2030) to extract the relative phase shift and amplitude. A further DAC output is used to control the qubit resonance frequency by providing flux bias to the transmon's DC-SQUID.\\
	}
	\label{fig:setup}
\end{figure*}

\begin{figure*}
	\includegraphics[scale=1.15]{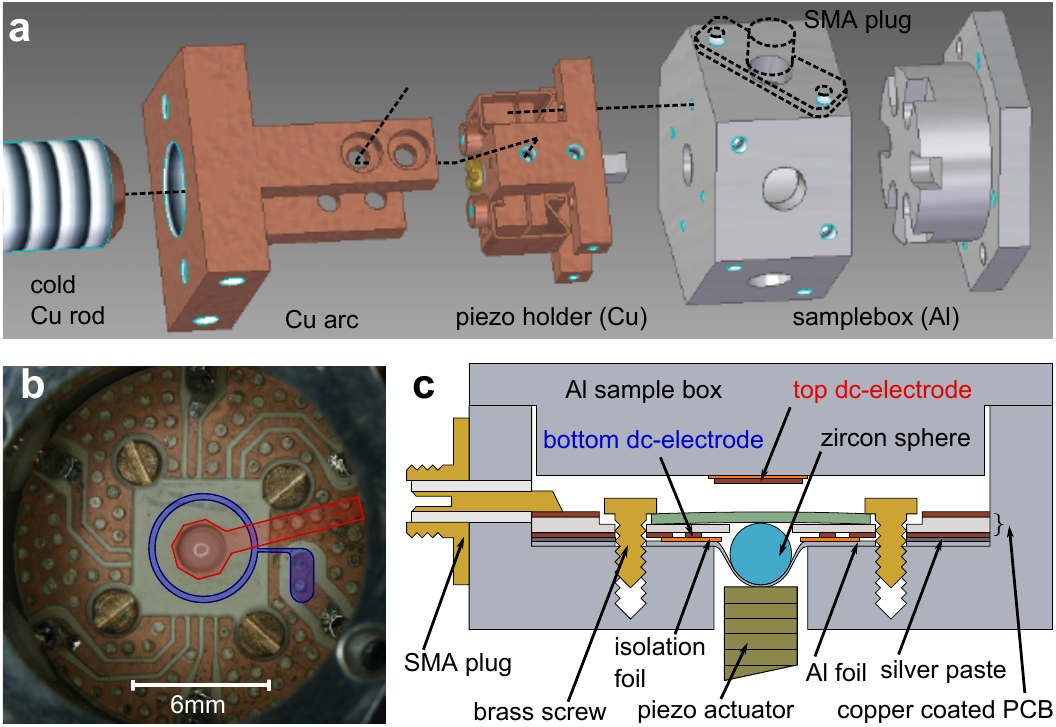}
	\caption{\textbf{Sample holder for strain- and E-field tuning of defects.} \textbf{a} Exploded view showing the aluminum sample box (right) and the piezo housing which are screwed to the cryostat's cold finger. \textbf{b} Photograph of the inside of the sample box, showing the printed circuit board (PCB) with coplanar microwave transmission lines and four screws that hold the chip (not installed here) against the pressure generated by the piezo. The red octagon depicts the top DC-electrode that is glued to the lid of the sample box. The blue ring indicates the bottom electrode that is milled from the backside metallization of the PCB. The qubits were located close to the center of the chip, where the electric field generated by the electrodes has its largest homogeneity.
	\textbf{c} Cross-section through the sample box. The zircon sphere is used to avoid shear stress on the piezo and to thermally isolate the piezo actuator and the sample chip. An aluminum foil between the sphere and the piezo serves for electromagnetic shielding.
	}
	\label{fig:sampleholder}
\end{figure*}

\begin{figure*}
	\includegraphics[width=\textwidth]{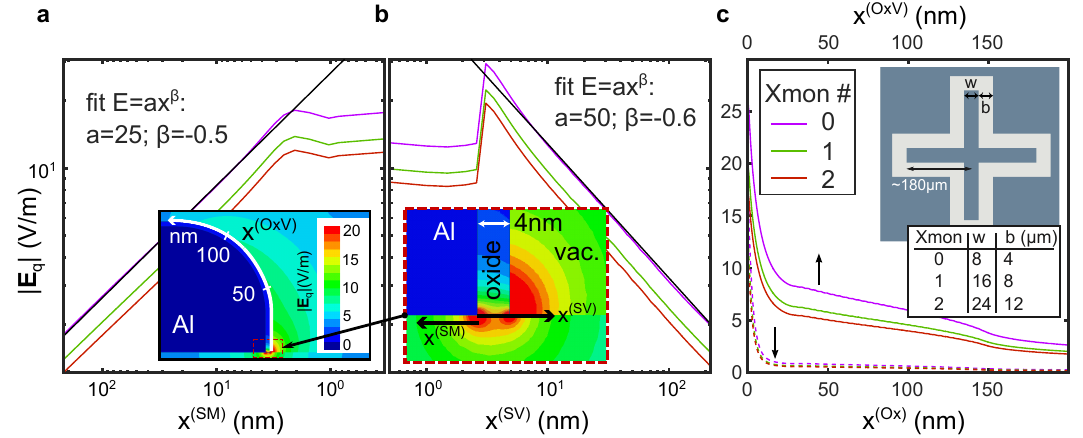}
	\caption{\textbf{Simulation of the electric field induced by the qubit plasma oscillations.} The electric potential of the qubit island is set to a DC-voltage of $U_\mathrm{rms} = 5 \mu$V which is about the root mean square of the oscillating voltage between the transmon electrodes. 
	\textbf{a,b} Double-logarithmic plot of $|E_q|$ for each qubit (see legend in \textbf{c}), at \textbf{a} the substrate-metal (SM) and \textbf{b} the substrate-vacuum (SV) interface, along the spatial axes as defined in the inset. 
	The fits (straight black lines) indicate a polynomial decay law of the field strength in agreement with results by Barends et al.~\cite{Barends13}. The nearly constant part at x $<$ 2 nm corresponds to the inside of the native surface oxide (having a thickness of 4 nm in the simulation). 
	\textbf{c} $|Eq|$ in the native oxide layer of the aluminum film (dashed lines, bottom axis) and at the interface of the native oxide to vacuum (solid lines, top axis). The inset illustrates the geometries of the Xmon samples used for simulations and actual measurements in this work.
	}
	\label{fig:simuqubit}
\end{figure*}

\begin{figure*}
	\includegraphics[width=8cm,height=6.5cm]{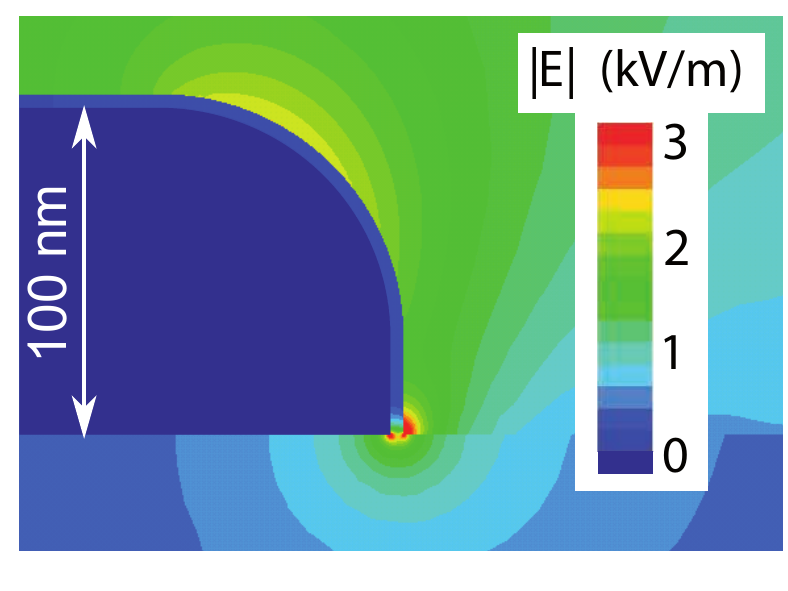}
	\caption{Simulation of the electric field near the qubit film edge when a voltage difference of 1V is applied between the top and bottom electrodes.
	}
	\label{fig:Efield_tb}
\end{figure*}

\begin{figure*}
	\includegraphics[width=18cm,height=7.5cm]{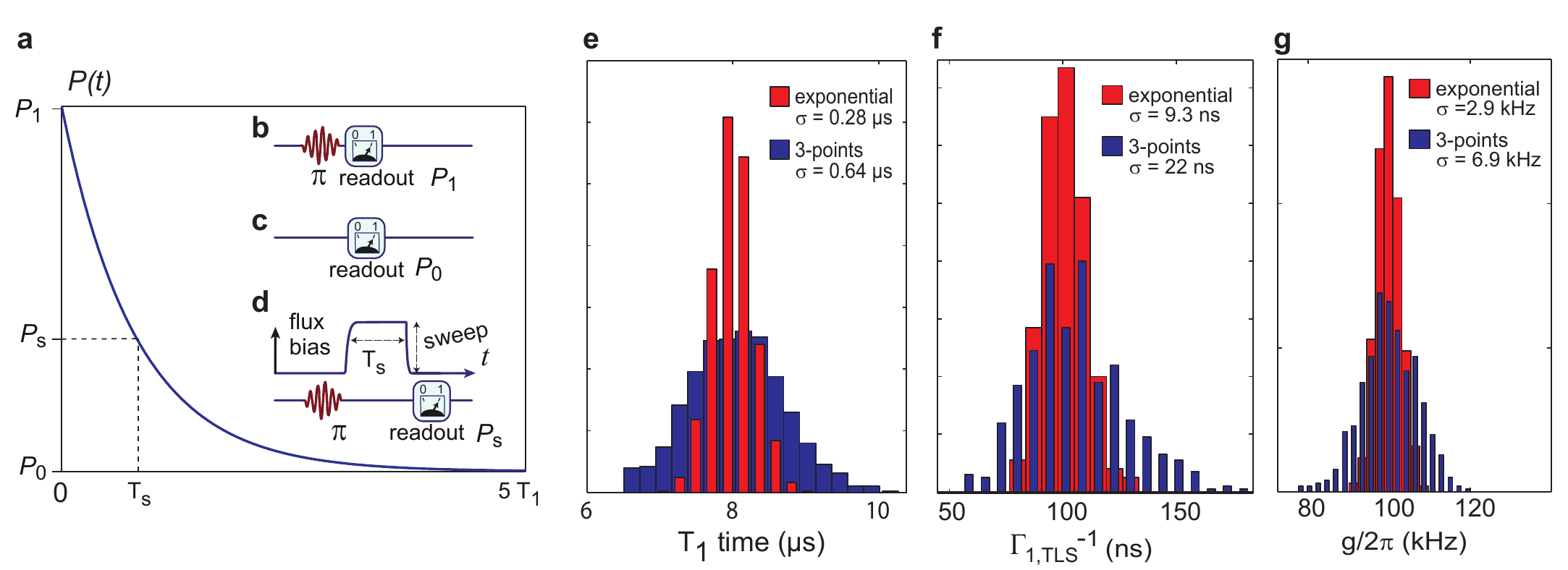}
	\caption{\textbf{Protocol to measure the frequency-dependence of the qubit's $T_1$ time using 3 data points.} \textbf{a} shows the exponentially decaying probability to find the qubit in its excited state for increasing delay time between qubit excitation by a resonant $\pi$-pulse and readout. The insets depict the used pulse sequences: \textbf{b} immediate readout after the $\pi$-pulse results in the maximum signal $P_1$ corresponding to the fully excited qubit. \textbf{c} No $\pi$-pulse is applied to measure the background signal $P_0$ corresponding to the qubit in its ground state. \textbf{d} A detuning flux pulse of duration $T_S\approx 0.5\,T_1$ and swept amplitude probes the loss of qubit population due to resonant TLS interaction and its frequency dependence. \textbf{e}-\textbf{g}: Results from numerical simulations to estimate the uncertainty of qubit and TLS parameters obtained from the 3-point method explained in \textbf{a}. \textbf{e} Distribution of qubit $T_1$ times obtained from noisy simulation data using full exponential fits (red) and the 3-point method (blue). \textbf{f} TLS decoherence times $\Gamma_{1,\mathrm{TLS}}^{-1}$ and \textbf{g} qubit-TLS coupling strength $g$ obtained from Lorentzian fits to the frequency-dependent energy relaxation rate of the qubit that was estimated from full exponential fits (red) and the 3-point method (blue), for a TLS with $\Gamma_{1,\mathrm{TLS}}^{-1}=100$ ns and $g/2\pi = 100$ KHz. The distribution's standard deviations $\sigma$ are quoted in the legends.
	}
	\label{fig:swapspec}
\end{figure*}

\clearpage 

\begin{figure*}
	\includegraphics[width=0.95\textwidth]{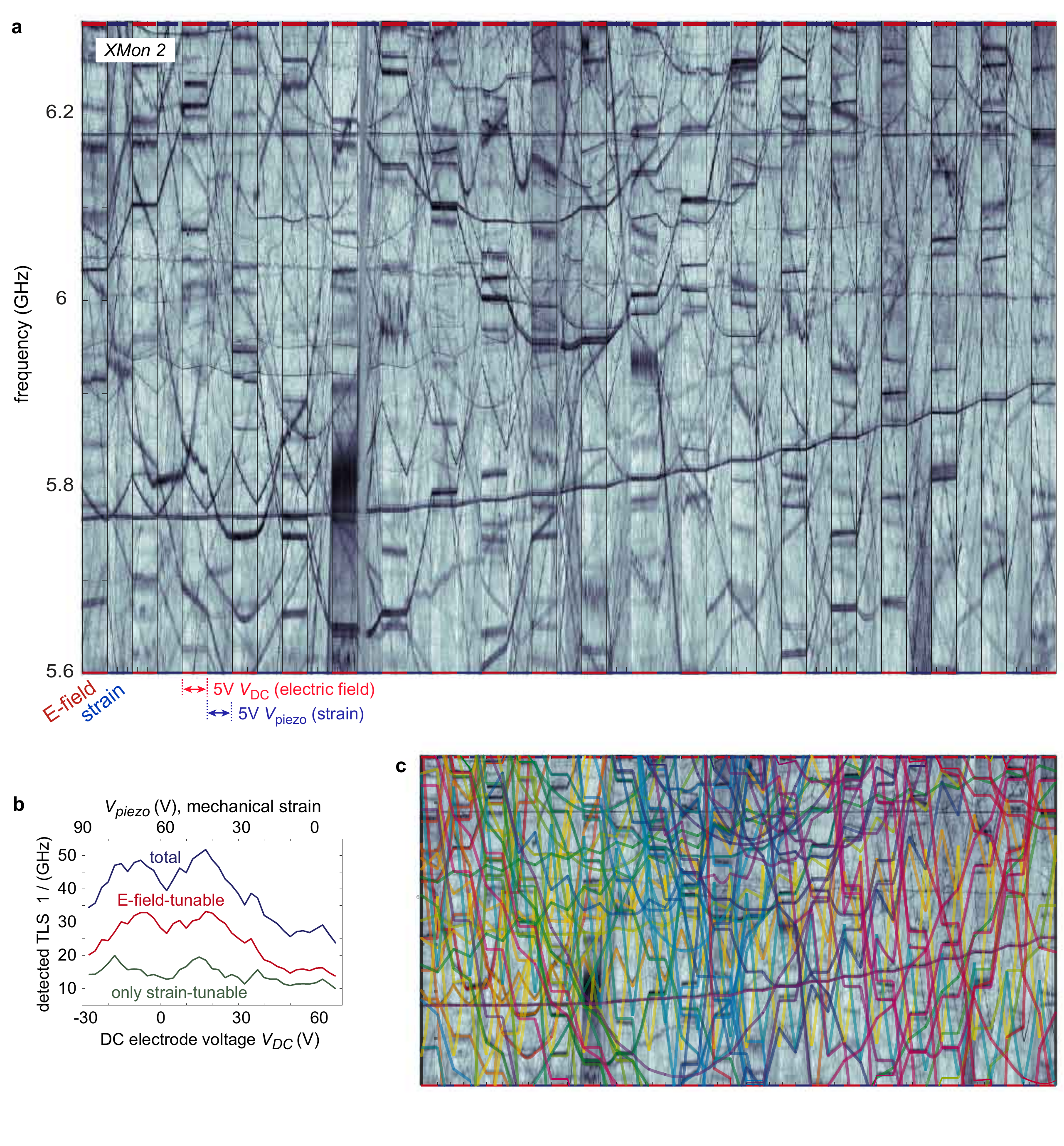}
	\caption{\textbf{Data obtained on XMon 2.} \textbf{a} Full dataset of a $T_1$ spectroscopy measurement in dependence of strain (blue margins) and electric field (red margins). \textbf{b} Number of detected TLS per GHz. The top and bottom axes indicate the applied voltages for the data shown in \textbf{a}. \textbf{c} Same data as in \textbf{a} with superimposed fits to the hyperbolic dependence of TLS resonance frequencies on strain and electric field.  The figure resolution has been reduced to limit the file size.
	} 
	\label{fig:Xmon2_supp}
\end{figure*}

\begin{figure*}
	\includegraphics[width=0.95\textwidth]{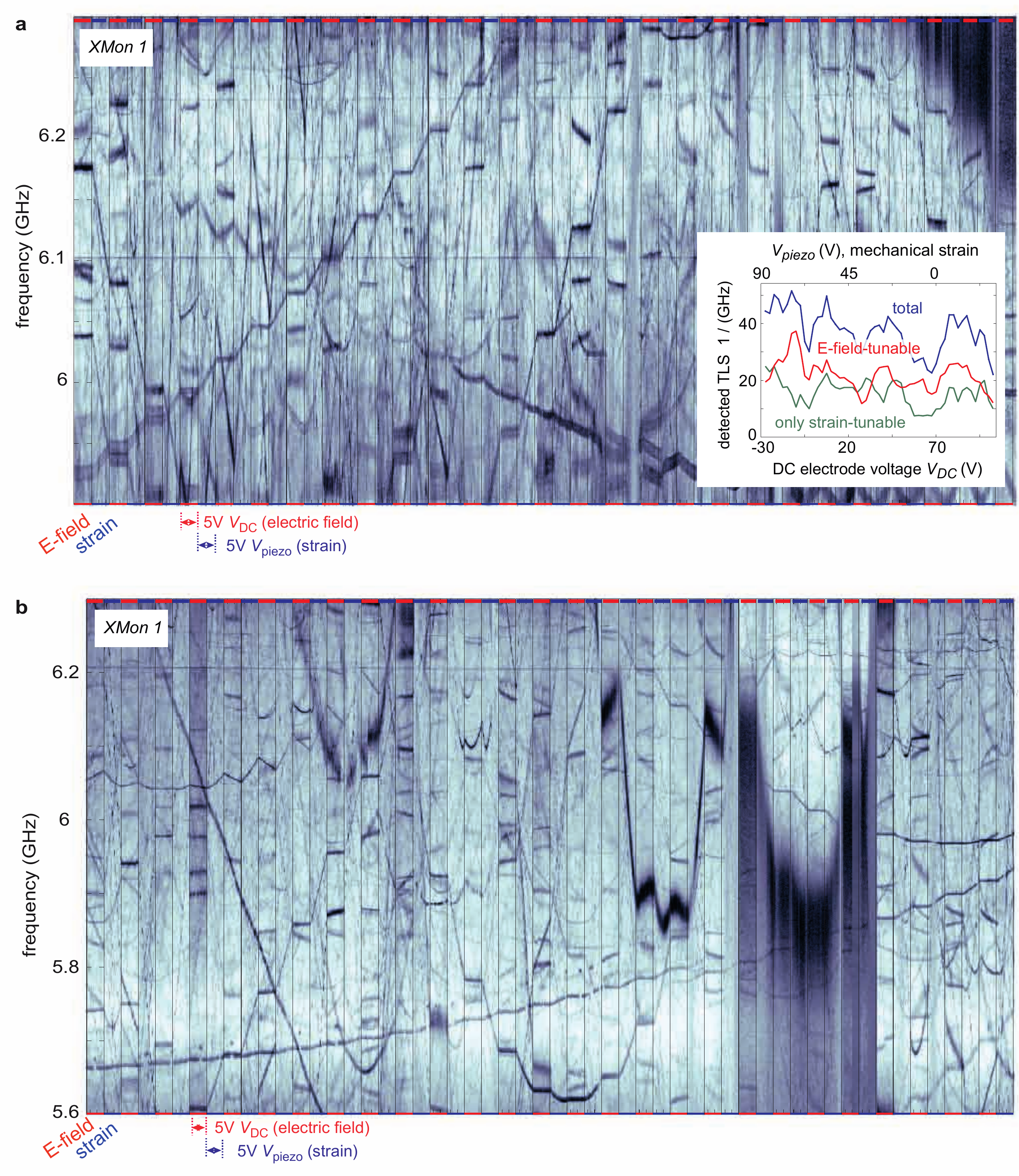}
	\caption{\textbf{Data obtained on XMon 1.} 
		\textbf{a} and \textbf{b} were obtained on the same qubit but in two different cool-downs. \textbf{b} High-resolution measurement for precise fitting of the TLS' Lorentzian resonance lineshape. The measurement duration was about 6 days. The figure resolution has been reduced to limit the file size.
	}  
	\label{fig:Xmon1_supp}
\end{figure*}

\begin{figure*}
	\includegraphics[width=0.8\textwidth]{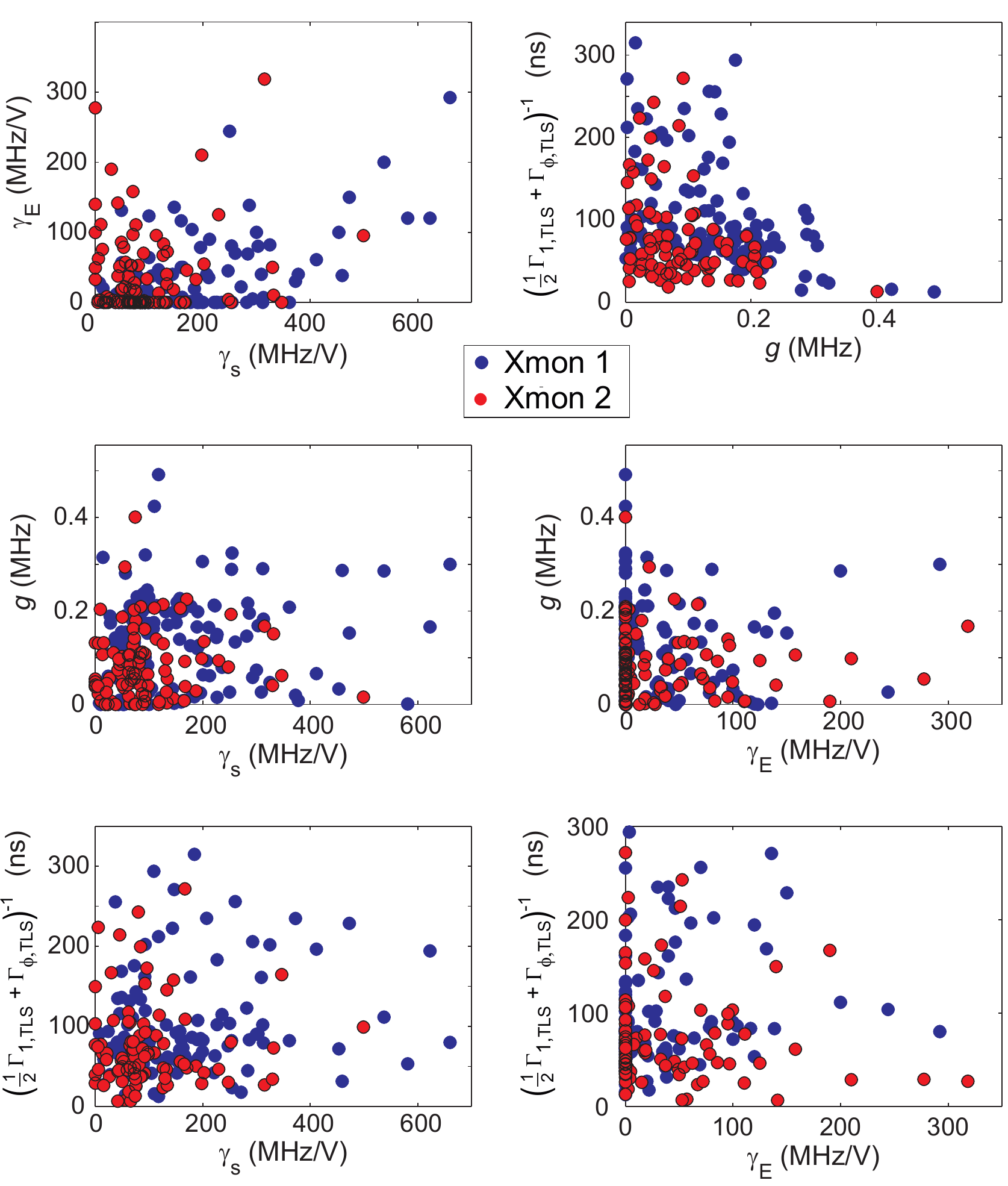}
	\caption{\textbf{Scatter plots of extracted TLS parameters.} $\gamma_E$ and $\gamma_S$ are the TLS' coupling strengths to the applied electric field and mechanical strain, respectively, which are obtained from fits to hyperbolic traces in data as shown in Fig.~3\textbf{a} of the main text. $g$ is the coupling strength between qubit and TLS and $(\Gamma_\mathrm{1,TLS}/2 + \Gamma_\mathrm{\Phi,TLS})^{-1}$ is the TLS' coherence time, which are extracted from Lorentzian fits to individual dips in the frequency-dependence of the qubit's energy relaxation rate as shown in Fig.~3\textbf{b}.
	} 
	\label{fig:allstats}
\end{figure*}

\end{document}